\documentclass[prl,twocolumn,superscriptaddress,showpacs]{revtex4}

\usepackage{bm}
\usepackage{latexsym}
\usepackage{dcolumn}
\usepackage{amsmath}
\usepackage{amssymb}

\newcommand{\iwr}{%
Zhou Pei-Yuan Center for Appl. Math., Tsinghua University,
Beijing 100084, China
}

\begin{document}
%

\title{An Onsager-like Relation for the Lattice Boltzmann Method}
\author{Wen-An Yong}
\email{wayong@tsinghua.edu.cn}
\affiliation{\iwr}

%
\begin{abstract}\noindent
An Onsager-like relation is proposed as a new criterion for
constructing and analysing the lattice Boltzmann (LB) method. For LB
models obeying the relation, we analyse their linearized stability,
establish their diffusive limit, and find new constraints for those
with free parameters. The new relation seems of fundamental
importance for the LB method.
\end{abstract}
\maketitle


The lattice Boltzmann (LB) method is an effective and viable tool for
simulating complex fluid flow problems. Historically, the method was
evolved from the lattice gas
automata \cite{FHP, Wo}. It is now well understood as a special
discretization of the Boltzmann equation \cite{HL1}. The kinetic
origin enables the method to naturally accomodate a variety of
boundary conditions for flows with complex geometry. Besides its
effectiveness and simplicity (see Eq.~(\ref{1}) below), the method
has a clear and solid physical interpretation \cite{CD, SKC}. These
advantages make it have vast applications. Indeed, the method has
been successfully used in a wide spectrum of areas including
turbulent flows, microflows, multi-phase and multi-component flows,
and particulate suspensions. It is becoming a serious alternative to
traditional computational methods in fluid dynamics. We refer to
\cite{CD, SKC, YMLS} for reviews of the method and its applications.

In spite of the vast and successful applications, the numerical
stability of the method has not been well understood but has
attracted much attention \cite{BYK, JY1, JY2, KCA, LL, SC, WMS, YL1,
YL2}. In this work, we intend to understand the LB method from the
viewpoint of nonequilibrium thermodynamics \cite{GM, KP}. This
understanding will provide a new insight into the method and a novel
approach to study the stability.

The general form of the LB method is
\begin{align}\label{1}
f_i({\bf x}_\mu + {\bf c}_i, t + 1) = f_i({\bf x}_\mu, t) + J_i({\bf f}({\bf x}_\mu, t))
\end{align}
for $i = 1, 2, \cdots, N$ and $\mu$ in a finite or countable set. Here
$f_i=f_i({\bf x}_\mu, t)$ is the probability of finding a fluid particle,
at site ${\bf x}_\mu\in{\mathbb R}^d$ and time $t\ge 0$, that travels with
velocity ${\bf c}_i\in{\mathbb R}^d$, $J_i({\bf f})$ is the $i$-th collision term,
and ${\bf f} = (f_1, f_2, \cdots, f_N)^T$ with the superscript $T$ denoting the
transpose operation. We will often write
$$
J({\bf f}) = (J_1({\bf f}), J_2({\bf f}), \cdots, J_N({\bf f}))^T.
$$

Eq.~(\ref{1}) is fixed by specifying (or constructing) the discrete
velocities ${\bf c}_i$ and collision terms $J_i({\bf f})$. This is
usually done by requiring (\ref{1}) to possess the following
properties: consistency with certain conservation laws, Galilean
invariance, isotropy, small or no compressible effects,
velocity-independent pressure, and so on. See \cite{CD, SKC, LL} for
details.

In this letter, we will propose a new requirement for the
construction. This requirement has its origin in non-equilibrium
thermodynamics. According to de Groot and Mazur \cite{GM},
nonequilibrium thermodynamics classifies irreversible phenomena into
three types: scalar, vectorial and tensorial processes. Typical
examples are chemical reaction, heat conduction and viscous flows,
respectively.

Our starting point is to understand that the $N$ equations in
(\ref{1}) describe $N$ scalar irreversible processes and the collision
terms $J_i({\bf f})$ are the corresponding irreversible fluxes.
Moreover, the thermodynamic forces are identified as the gradient
of a possibly-existing entropy-density function $H(f)$ \cite{Y4}.
With such identifications, the Onsager reciprocal relation
reads as
\begin{equation}\label{2}
\begin{split}
J({\bf f})=S({\bf f})H_f({\bf f}), & \qquad S({\bf f})=S({\bf f})^T,
\\[2mm]
\mbox{The null space of matrix } & S({\bf f}) \ \mbox{is independent
of} \ {\bf f}.
\end{split}
\end{equation}
Here $H_f({\bf f})$ stands for the gradient of $H=H({\bf f})$ with
respect to $\bf f$, that is,
$$
H_f({\bf f}) : = \left({\partial_{f_1}}, {\partial_{f_2}}, \cdots,
{\partial_{f_N}}\right)^TH({\bf f}).
$$
Let us mention that the gradient of the entropy-density function
$H=H(f)$ has been used in \cite{AK, KCA} to construct entropic LB
methods.

Remark that, in the classical Onsager relation \cite{GM}, the
symmetric matrix $S=S({\bf f})$ is constant. Recently in \cite{Y4},
the author looses the constancy requirement and proposes (2) for
general scalar processes. The new independence in (2) expresses the
fact that physical laws of conservation hold true, no matter what
state the underlying thermodynamical system is in (equilibrium,
non-equilibrium, and so on). In addition, the second law of
thermodynamics requires that the symmetric matrix $S({\bf f})$ have
a sign (non-positive).

If an LB model admits the relation (\ref{2}), one can easily show
that it allows an H theorem with $H({\bf f})$ as the entropy-density
function (see \cite{Y4, KCA}). Therefore, the relation (\ref{2}) is
not realistic, for many widely used LB models were shown in
\cite{YL1, YL2} not to admit an H theorem. On the other hand, it
follows from (2) that $S({\bf f}_*)H_f({\bf f}_*) = J({\bf f}_*) =0$
for $\bf f_*$ satisfying $J(\bf f_*)=0$ (in equilibrium). Thus,
$H_f({\bf f}_*)$ is in the null space of $S({\bf f}_*)$ and thereby
in that of $S({\bf f})$, since the null space of $S({\bf f})$ is
independent of $\bf f$. Therefore, we have
$$
S({\bf f})H_f({\bf f}_*) = 0
$$
for any $\bf f$. Now we differentiate the two sides of Eq.~(\ref{2})
with respect to $\bf f$ and compute at $\bf f_*$ to obtain
\begin{equation}\label{3}
\begin{split}
J_f({\bf f}_*)= & S({\bf f}_*)H_{ff}({\bf f}_*) +
\frac{\partial}{\partial f}[S({\bf f})H_{f}({\bf f}_*)]|_{\bf f=\bf
f_*}\\
= & S({\bf f}_*)H_{ff}({\bf f}_*) .
\end{split}
\end{equation}
Here $H_{ff}({\bf f})$ is the Hessian of the entropy function
$H({\bf f})$.

From Lemma 2.1 in \cite{YL2} we know that, if the discrete-velocity
set is such that ${\bf c}_i\ne{\bf c}_j$ whenever $i\ne j$, the
possibly-existing entropy-density functions $H({\bf f})$ for (\ref{1})
must be of the following form
$$
H({\bf f})=\sum_{i=1}^Nh_i(f_i).
$$
That is, $H({\bf f})$ does not contain any cross-terms. Thus,
the Hessian $H_{ff}({\bf f})$ must be a diagonal and positive-definite
matrix. The positive-definiteness follows from the strict convexity of
$H({\bf f})$ as an entropy-densitiy function.

Inspired by these considerations, we propose the following notion. The
LB method (\ref{1}) is said to obey an {\it Onsager-like relation}
at ${\bf f}={\bf f}_\ast$ satisfying $J({\bf f}_\ast)=0$,
if the Jacobian matrix of $J({\bf f})$
evaluated at ${\bf f}={\bf f}_\ast$ can be decomposed as
\begin{align}\label{4}
J_f({\bf f}_\ast) = S({\bf f}_\ast)D({\bf f}_\ast),
\end{align}
where $S({\bf f}_\ast)$ is a symmetric non-positive-definite matrix and
$D({\bf f}_*)$ is a diagonal positive-definite matrix.

In \cite{JY1, BYK}, we verified that many LB models admit the
following structure: there is an invertible $N\times N$-matrix
$P$ such that $P^TP$ is diagonal and
\begin{align}\label{5}
PJ_f({\bf f}_\ast)=
-\mbox{diag}(\lambda_1, \lambda_2, \cdots, \lambda_N)P
\end{align}
with the $\lambda_i$'s non-negative. In this situation, we have
$$
J_f({\bf f}_\ast) = -P^{-1}
\mbox{diag}(\lambda_1, \lambda_2, \cdots, \lambda_N)P^{-T}P^TP.
$$
This is the relation (\ref{4}), because $P^TP$ is diagonal and
$P^{-1}\mbox{diag}(\lambda_1, \lambda_2, \cdots, \lambda_N)P^{-T}$
is symmetric.

It is interesting to note that the relation (\ref{4}) is equivalent to (\ref{5}).
To see this, we set $\Lambda = \sqrt{D({\bf f}_\ast)}$. As
$\Lambda S({\bf f}_\ast)\Lambda$ is symmetric and non-positive definite,
there is an orthogonal matrix $U$ such that
$$
\Lambda S({\bf f}_\ast)\Lambda
=-U^T\mbox{diag}(\lambda_1, \lambda_2,\cdots, \lambda_N)U
$$
and the $\lambda_k$'s are non-negative. Thus, it follows from Eq. (\ref{4}) that
$$
J_f({\bf f}_\ast)= S({\bf f}_\ast)\Lambda^2= -\Lambda^{-1}U^T
\mbox{diag}(\lambda_1, \lambda_2,\cdots, \lambda_N)U\Lambda.
$$
Set $P=U\Lambda$ and notice that $U^{-1}=U^T$. Then
$P^TP=\Lambda^2=D({\bf f}_\ast)$ is diagonal and $PJ_f({\bf f}_\ast)=
-\mbox{diag}(\lambda_1, \lambda_2,\cdots, \lambda_N)P$. Hence the relation (\ref{4}) also
implies (\ref{5}).

Generally speaking, the invertible matrix $P$ in (\ref{5}) differs from the
transformation matrix $M$ used in constructing the multiple-relaxation-time
LB method \cite{dH, LL}. The latter is such that $MM^T$ is diagonal.

In the rest of this article, we present some consequences of
the Onsager-like relation (\ref{4}) or its equivalent version (\ref{5}).

First, the relation (\ref{4}) can be used as an analytic criterion
to fix LB models with free parameters. In \cite{BYK}, we require
that a number of existing LB models with free parameters obey the
equivalent relation (\ref{5}) and find new constraints for the free
parameters. In this way, either the free parameters are fixed or
their freedom is reduced considerably. The results coincide with
those obtained through numerical tests and/or guesswork.

An example is the D2Q9 model with two parameters $\alpha$ and $\beta$ \cite{SC}.
Here $d=2, N=9$, the discrete velocities are ${\bf c}_9 = 0$,
\begin{align*}
\{{\bf c}_i: 1\leq i\leq4\}=  & \{(\pm1, 0)^T, (0, \pm1)^T\},\\
\{{\bf c}_i: 5\leq i\leq8\}= & \{(\pm1, \pm1)^T\},
\end{align*}
and the collsion terms are
$$
J_i({\bf f}) = \frac{f^{eq}_i(n, {\bf v}) - f_i}{\tau}
$$
with $\tau$ a relaxation time. Furthermore,
\begin{align*}
f^{eq}_9(n, {\bf v}) = & \alpha n -\frac{2}{3}n|{\bf v}|^2, \\
f^{eq}_i(n, {\bf v}) = & \beta n + \frac{n}{3} {\bf v}{\bf c}_i +
\frac{n}{2}({\bf v}{\bf c}_i)^2 -\frac{n}{6}|{\bf v}|^2
\end{align*}
for $i=1, 2, 3, 4$,
$$
f^{eq}_i(n, {\bf v}) = \frac{(1 - \alpha - 4\beta)}{4}n +
\frac{n}{12}{\bf v}{\bf c}_i +
\frac{n}{8}({\bf v}{\bf c}_i)^2 -\frac{n}{24}|{\bf v}|^2
$$
for $i=5, 6, 7, 8$, and
$$
n = \sum_{i = 1}^9 f_i, \qquad n{\bf v} = \sum_{i = 1}^9 {\bf c}_i f_i.
$$

In \cite{BYK}, we verified that the above parametrized D2Q9 model
obeys the equivalent version (\ref{5}) if
\begin{equation}\label{6}
\alpha\in(0, 1)\qquad \mbox{and}\qquad \alpha + 5\beta = 1.
\end{equation}
This is satisfied by both $(\alpha, \beta) = (4/9, 1/9)$ used in the well-known
D2Q9 model \cite{QDL} and $(\alpha, \beta) = (2/7, 1/7)$ used in \cite{Sk}.
Moreover, the numerical tests in \cite{BYK} support the other choices of
$(\alpha, \beta)$ satisfying condition (\ref{6}).

Next, we show that the equivalent version (\ref{5}) provides a convenient setting
to analyse the linearized stability of the LB method (\ref{1}). Let ${\bf f}_\ast$
be a uniform equilibrium state and $\tilde{\bf f}$ the fluctuation. The linearized
LB method is
\begin{equation}\label{7}
\tilde {\bf f}_\mu(t + 1)=\tilde {\bf f}({\bf x}_\mu, t) +
J_{f}({\bf f}_\ast)\tilde {\bf f}({\bf x}_\mu, t) ,
\end{equation}
where
$$
\tilde{\bf f}_\mu(t)=(\tilde f_1({\bf x}_\mu + {\bf c}_1, t),
\cdots, \tilde f_N({\bf x}_\mu + {\bf c}_N, t))^T.
$$
For the sake of simplicity, we consider only the periodic initial data for
(\ref{7}), where $\mu$ ranges in a finite set. For other cases, see \cite{JY2}.

Assume the Onsager-like relation (\ref{4}) holds at the equilibrium state
${\bf f}_\ast$. Then we have the equivalent relation (\ref{5}). Multiplying the
linearized equation (\ref{7}) with $P$ from the left gives
$$
P\tilde{\bf f}_\mu(t + 1)=
\mbox{diag}(1- \lambda_1, 1- \lambda_2, \cdots, 1- \lambda_N)
P\tilde{\bf f}({\bf x}_\mu, t).
$$
Thus, if
\begin{equation}\label{8}
\lambda_i\in[0, 2]\qquad \forall \ i,
\end{equation}
it is obvious that
\begin{equation}\label{9}
|P\tilde {\bf f}_\mu(t + 1)|^2 \leq |P\tilde{\bf f}(x_\mu, t)|^2 .
\end{equation}
Here $|\bf f|$ is the Euclidean length of the $N$-vector $\bf f$.
On the other hand, since $P^TP$ is a diagonal matrix, say diag$(a_1, a_2, \cdots, a_N)$,
it follows that
\begin{align*}
|P\tilde {\bf f}_\mu(t + 1)|^2
= & \tilde {\bf f}_\mu(t + 1)^TP^TP\tilde {\bf f}_\mu(t + 1)\\
= & \sum_{i=1}^Na_i{\tilde f}_i^2({\bf x}_\mu + {\bf c}_i, t + 1).
\end{align*}
Because ${\bf x}_\mu + {\bf c}_i$ is a lattice node, we deduce that
\begin{align*}
\sum_\mu|P\tilde {\bf f}_\mu(t + 1)|^2 = & \sum_\mu\sum_{i=1}^Na_i{\tilde f}_i^2({\bf x}_\mu + {\bf c}_i, t + 1)\\
= & \sum_\mu\sum_{i=1}^Na_i{\tilde f}_i^2({\bf x}_\mu, t + 1)\\
= & \sum_\mu|P{\tilde {\bf f}}({\bf x}_\mu, t + 1)|^2.
\end{align*}
Now, summing up Eq. (\ref{9}) over all $\mu$ (in the finite set!) gives the following
inequality
\begin{align}\label{10}
\sum_\mu|P\tilde {\bf f}(x_\mu, t +1)|^2\leq
\sum_\mu|P\tilde{\bf f}(x_\mu, t)|^2.
\end{align}
This simply means the stability of the linearized LB method (\ref{7}) under the condition
(\ref{8}). Note that for the D2Q9 model above, (\ref{8}) is nothing but the well-known
condition $\tau\geq 1/2$.

Our derivation of the inequality (\ref{10}) does not involve the von
Neumann stability analysis, which was used in the previous works
\cite{SC, WMS, LL} on the stability of the LB method. Moreover, the
derivation is different from those in \cite{KCA}, does not involve
any entropy-density function, and works for LB models violating H
theorems. Finally, our approach can be easily extended to other
cases, even with boundaries \cite{JY2}.

Finally, we mention that the Onsager-like relation (\ref{4}) ensures the validity
of the formal diffusive limit for the continuous version of the LB model (\ref{1})
parametrized with $\epsilon >0$:
\begin{equation}\label{11}
\partial_t f_i + \frac{1}{\epsilon}{\bf c}_i\cdot\nabla_x f_i =
\frac{1}{\epsilon^2}J_i(\bf f) .
\end{equation}
It is well-known (see, e.g., \cite{SC, JY1}) that this system of parametrized
partial differential equations is related closely to the LB method (\ref{1}).
Suppose the discrete-velocity set is symmetric in the sense that
$$
\{{\bf c}_i: i=1, 2, \cdots, N\}=\{-{\bf c}_i: i=1, 2, \cdots, N\},
$$
the parametrized model (\ref{11}) is consistent with the incompressible Navier-Stokes
equaton and obeys the equivalent version (\ref{5}) at a quiescent state ${\bf f}_\ast$.
We rigorously verified in \cite{JY1} that the solution $f_i^\epsilon({\bf x}, t)$ to (\ref{11}) with prepared initial data satisfies
\begin{equation}\label{12}
\sum_{i=1}^Nf_i^\epsilon({\bf x}, t){\bf c}_i = \epsilon{\bf v}({\bf x}, t) + O(\epsilon^3)
\end{equation}
as $\epsilon$ goes to zero. Here ${\bf v}({\bf x}, t)$ is the velocity of the fluid under
consideration. The details are given in \cite{JY1}.

In conclusion, we have proposed an Onsager-like relation (\ref{4})
and its equivalent version (\ref{5}) as a new requirement for
analysing and constructing LB models. For LB models obeying the
Onsager-like relation, we analyse their linearized stability,
established their diffusive limit, and found new constraints for
those with free parameters. On the basis of these consequences, we
believe that the new relation (\ref{4}) is of fundamental importance
for the LB method. Finally, we expect that the Onsager-like relation
(\ref{4}) can be used as a guide to construct stable LB models.

\end{document}